**Comment on "Resonant X-ray diffraction studies on the charge ordering in magnetite".**

In a recent letter, E. Nazarenko et al.[1] report a resonant X-ray diffraction (RXD) study in the low temperature phase of magnetite. The paper puts forward the quantitative determination of an effective charge ordering (CO) of 0.24 electron among the octahedral iron atoms in the insulating phase. The authors have corrected the paper in a subsequent erratum[2]. The ordering scheme now coincides with that proposed by Wright et al [3]. They argument that their results are supported by fitting of the energy dependence of 32 independent reflections in resonant condition.

From these 32 reflections, they report on eight as representative, showing the calculated spectra with and without CO. Among them, the two reflections of the type (*-4 4 odd/2*) are claimed in Nazarenko et al. as sensitive to the $Fe_3$- $Fe_4$ ordering, i.e the structure factors must contain the difference between the atomic scattering factors $f_{Fe3}$ and $f_{Fe4}$. We have calculated these structure factors, the terms of it regarding the octahedral iron atoms within the Wright's description[2] are: $F(-4\ 4\ 3/2) = 0.956\ f_{Fe2} - 1.201\ f_{Fe3} - 0.146\ f_{Fe4}$ and $F(-4\ 4\ 5/2) = -0.956\ f_{Fe2} + 0.911\ f_{Fe3} + 0.135\ f_{Fe4}$. Accordingly, both reflections are essentially proportional to the differences between $f_{Fe2}$ and $f_{Fe3}$ instead of $f_{Fe3}$ and $f_{Fe4}$. In the proposed ordering, $Fe_2$ and $Fe_3$ have the same charge. Therefore, no difference between CO and non-CO should be obtained in the theoretical calculations (Fig. 4 in ref 1). The authors must account for this strong incompatibility.

The incoherence stated above together with the fact that the charge disproportion supposedly found by the authors is very small shows the absence of ionic ordering in magnetite below the Verwey transition temperature ($T_V$). We already showed that the charge disproportion must be less than 0.25 electron[4] for any kind of charge ordering and 0.1 electron (corresponding to 0.2 electron for $Fe_1$ and $Fe_2$ in the Wright et al. ordering) along the c-axis.[5] Nazarenko et al claim for a charge disproportion just within these limits established by our RXD works[4-6] so similar conclusions regarding the electronic state of the octahedral iron atoms should be derived from the two studies. Indeed, Nazarenko et al confirms that the octahedral iron atoms are in a intermediate valence state so they are far from the ionic states as Verwey proposed[7] (non-integer charge localization is not considered in the Verwey model) and the remaining electron is delocalized among different iron atoms also below Tv. Therefore, this very small charge disproportion does not justify the Nazarenko et al. claim for an exclusive electronic origin of the Verwey transition.

Joaquin Garcia*, Gloria Subías*, Javier Blasco*, Ma Grazia Proietti* and Hubert Renevier#

* ICMA, CSIC- Universidad de Zaragoza. 50009 Zaragoza, Spain

# CEA, Dep. de Rech. Fondamentale sur la Mat. Condensée, SP2M/NRS, 38054 Grenoble Cedex 9, France